\documentclass{aastex701}

\usepackage{siunitx}
\usepackage{amsmath}
\DeclareSIUnit{\erg}{erg}
\DeclareSIUnit{\ergs}{ergs}
\journalinfo{}

\begin{document}

\title{The Wonderful World of Binary Stars\footnote{Based on data from ESO Prog. ID 60.A-9501(D)}}

\author[orcid=0009-0001-5140-8220,gname=Andrea,sname=Barone]{Andrea Barone}
\affiliation{Stockholm University, Astronomy Department, AlbaNova University Center SE-106 91 Stockholm, Sweden}
\email{andrea.barone@astro.su.se}  

\author[orcid=/0000-0002-9486-4840, gname=Henri, sname=M.J.Boffin]{Henri M. J. Boffin} \affiliation{European Southern Observatory, Karl-Schwarzschild-Strasse 2,
85748, Garching bei München,
Germany} \email{hboffin@eso.org}

\author[orcid=0009-0001-3865-0119,gname=Beatrice, sname='Caccherano']{Beatrice Caccherano} 
\affiliation{Queen Mary University of London, Mile End Road, London E1 4NS, UK}
\email{b.caccherano@qmul.ac.uk}

\author[orcid=0009-0009-6673-0851,gname=Simona, sname='Di Stefano']{Simona Di Stefano}
\affiliation{Dipartimento di Fisica, Sezione di Astronomia, Università di Trieste, Via Tiepolo 11, I-34143 Trieste, Italy}
\affiliation{INAF, Osservatorio Astronomico di Trieste, Via Tiepolo 11, I-34143 Trieste, Italy}
\email{simona.distefano@phd.units.it}

\author[0009-0008-1918-5646]{Akhila Divakaran}
\affil{Instituto de Astronomía y Ciencias Planetarias, 
Universidad de Atacama, Copiapó, Atacama, Chile.} 
\email{akhila.divakaran.25@alumnos.uda.cl} 

\author[orcid=0009-0004-3114-2527,gname=Alexandra,sname=Stockwell Murphy]{Alexandra S. Murphy}
\affiliation{Division of Astrophysics, Department of Physics, Lund University, Box
118, 221 00, Lund, Sweden}
\email{al5517mu-s@student.lu.se}

\author[orcid=0000-0003-4009-8316, gname=María, sname=José Rain]{María José Rain} \affiliation{European Southern Observatory, Alonso de Córdova 3107, Vitacura, Región Metropolitana, Chile} \email{mrainsep@eso.org}

\author[orcid=0000-0002-7444-5315]{Elyar Sedaghati} 
\affiliation{European Southern Observatory, Alonso de Córdova 3107, Vitacura, Región Metropolitana, Chile}
\email{esedagha@eso.org}

\author[orcid=0009-0001-4452-4944]{Paul V. Steimle}
\affiliation{Max-Planck-Institut für Astronomie, Königstuhl 17, 69117 Heidelberg, Germany}
\email{pasteimle@mpia.de}  

\begin{abstract}
During the 2026 ESO La Silla Observing school, about twenty students attended lectures and performed observations to learn various aspects of observational astronomy. The school, scientifically organised by Elyar Sedaghati, took place during the first two weeks of February 2026, starting with talks and lectures at ESO Vitacura offices in Santiago, then continuing with four nights of observations at the observatory, using EFOSC2/NTT, HARPS+NIRPS/3.6m, culminating with three final days in Vitacura where the various groups analysed the obtained data and presented the results of their projects. The students were split into four groups and group 3, nicknamed Unicorns and supervised by Henri Boffin and Maria Jose Rain, was devoted to the study of binary stars. Several projects were considered and followed up by some of the six students in this group. The first subgroup used HARPS to study the Rossiter-McLaughlin effect in binary stars to infer the relative inclination of the rotation axis of the primary with respect to the orbital plane. A detailed study of the contact binary system HD 115264 led to the conclusion that the primary is well aligned, likely as a result of strong tidal forces within the binary. The second subgroup analysed blue straggler stars (BSS) in open clusters, using both HARPS and EFOSC2. With HARPS, they looked at some well-known long-period binary with the aim of determining their chemical abundances, thereby confirming their membership to the cluster, as well as looking for any chemical anomalies that might be explained by mass transfer. EFOSC2 was used to derive radial velocities of rapidly varying BSS. For one of them -- the star Rediet -- the students clearly detected and analysed the radial velocity variations due to the second overtone pulsation, thereby confirming its delta Scuti character. Finally, one student used EFOSC2 to study planetary nebulae (PN) -- taking nice images of some of these intricate objects, as well as doing time-resolved photometry and spectra of some others. In one case, the binary nature of the central star of the PN was proven, confirming some previous estimates done with ZTF. Each subgroup thus was able to obtain useful research results, which we present hereafter. 
\end{abstract}

\keywords{{stars: abundances} --- {blue stragglers} --- {techniques: spectroscopic} --- {Planetary nebulae} --- {Binary stars} --- {Close binary stars} -- {Spectroscopy} --- 
Radial velocity --- Delta Scuti variable stars  --- Stellar oscillations}

\pagebreak

\section{The Primary Star of the Eclipsing Binary HD 115264 is Aligned}

The Rossiter-McLaughlin (RM) effect \citep{Rossiter1924, McLaughlin1924} is a rapid shift in the radial velocity measurements of a rotating star caused by the passage of a transiting (or eclipsing) object. During transit, the object sequentially covers different sections of the star, causing a redshift when it blocks the stellar hemisphere rotating towards us, and vice versa. This effect is a deviation from the Doppler reflex induced by the gravitational interaction, and modelling it allows the derivation of the size ratio between the object and its host star, the stellar rotational speed, the impact parameter, and the sky-projected spin–orbit angle.

The RM effect helps probe binary formation, evolution, and tidal angular momentum exchange \citep[e.g.,][]{Winn2010, Triaud2011, Albrecht2012}. Many binary systems are unequal pairs with secondaries of masses $\leq 0.5~M_{\odot}$, comparable in size and temperature to hot-jupiters. This makes them ideal comparison samples with close-in giant planetary systems \citep{Triaud2014, Triaud2017}, which have been extensively studied with the RM effect since the early 2000s \citep[e.g.,][]{Queloz2000,Johnson2009,Lendl2014,2025AJ....170..274Z}.

We present a study of the RM effect of the close binary system HD 115264, composed of a bright (V=9.996) star of spectral type F3, mass $M_{\star,1} = 1.44 \pm 0.05\ M_{\odot}$ and radius $R_{\star,1}=1.48\pm0.02\ R_{\odot}$, and a companion of undefined spectral type, mass $M_{\star,2} = 0.32 \pm 0.01\ M_{\odot}$ and radius $R_{\star,2}=0.77 \pm 0.01\ R_{\odot}$ \citep{Guo2025}. The secondary has evolved off the main sequence and must have transferred mass to the primary, leading to a contact system. To confirm the mass of HD 115264 B and study the system's RM effect, we carried out radial velocity (RV) observations of the star using the  High Accuracy Radial velocity Planet Searcher \citep[HARPS;][]{Mayor2003} spectrograph, mounted on the European Southern Observatory's (ESO) 3.6 m telescope, in La Silla, Chile.

\subsection{Observations and data reduction} \label{sec: Observations and data reduction}
\subsubsection{\textit{TESS Photometry}}
HD 115264 (TIC~453784977) was observed by the TESS mission in sectors 11, 37, and 64. It is listed in the TESS Eclipsing Binary Catalog, with orbital period of $P = 0.409722 \pm 0.000009$~days; $T_{\rm{eff}}$ of $6661\pm249$~\rm{K} and log~\textit{g} of $4.2\pm0.1$~dex \citep{TESSEB2022}. We use the ephemerides derived from the TESS photometry to predict the transit window during the HARPS observations.

\subsubsection{\textit{HARPS Spectroscopy}}
Our RV observations consist of 15 high-resolution ($R \approx 115,000$) spectra of the system acquired on February 9th 2026 (during the La Silla Summer School 2026 observing programme), with fibre A on target and fibre B on the sky. The exposure time was set to 450 s, and the acquisition mode to EGGS, which resulted in a median signal-to-noise ratio $(S/N)$ of $\sim50$ per pixel at 550 nm. 

The spectra were calibrated and reduced using the dedicated HARPS data reduction software \citep[DRS;][]{Lovis2007}. The cross-correlation function (CCF) obtained from the HARPS DRS did not provide accurate RV measurements, possibly because of the low S/N. Instead, we compared with a theoretical F3 V stellar template from PHOENIX stellar atmosphere models \citep{Hauschildt1999} to search for prominent absorption features. We identified the Balmer series (i.e., H$_{\alpha}$, H$_{\beta}$, H$_{\gamma}$, H$_{\delta}$), and the calcium doublet (i.e., Ca {\sc II} \ 3933.66 Å and Ca {\sc II} \ 3968.47 Å), and fit a Lorentzian to the lines. We then extracted the wavelength corresponding to the line's peak and computed the RV shift using the Doppler formula. This alternative method increased our systematic uncertainties, since the Balmer lines are broadened, possibly leading to imprecision in the RV extraction, but allowed us to retrieve the velocity measurements despite limitations.

\subsection{Results} \label{sec: Results}
For each observation, we derived six RV measurements, one for each line analysed. We considered the median of each line-derived data point in the time series as our estimate. The error associated with each observation is the standard deviation of the RVs. 

From the TESS lightcurves and the radius estimates of \cite{Guo2025}, we derived a transit depth of $dF=0.27\pm 0.01$. Furthermore, we identified the transit window from the TESS ephemerides and masked all data points occurring during transit. We then fit a Keplerian model with a fixed period to the masked RV dataset to isolate the signal induced by the gravitational interaction of the two stars. From this fit, we obtained a semi-amplitude of $K = 38 \pm 2\ \text{km s}^{-1}$, which we used to infer a companion mass $M_{\star,2} = 0.20 \pm 0.03 \ M_{\odot}$. The mass we derived is not in agreement with the value of \cite{Guo2025}; however, our results are preliminary since the mass was inferred with too few out-of-transit points. Furthermore, \cite{Guo2025} derived the system's fundamental parameters by adopting an empirical law. More out-of-transit points would be needed to confirm the mass of the secondary.  

\begin{figure}
    \centering
    \includegraphics[width=0.7\linewidth]{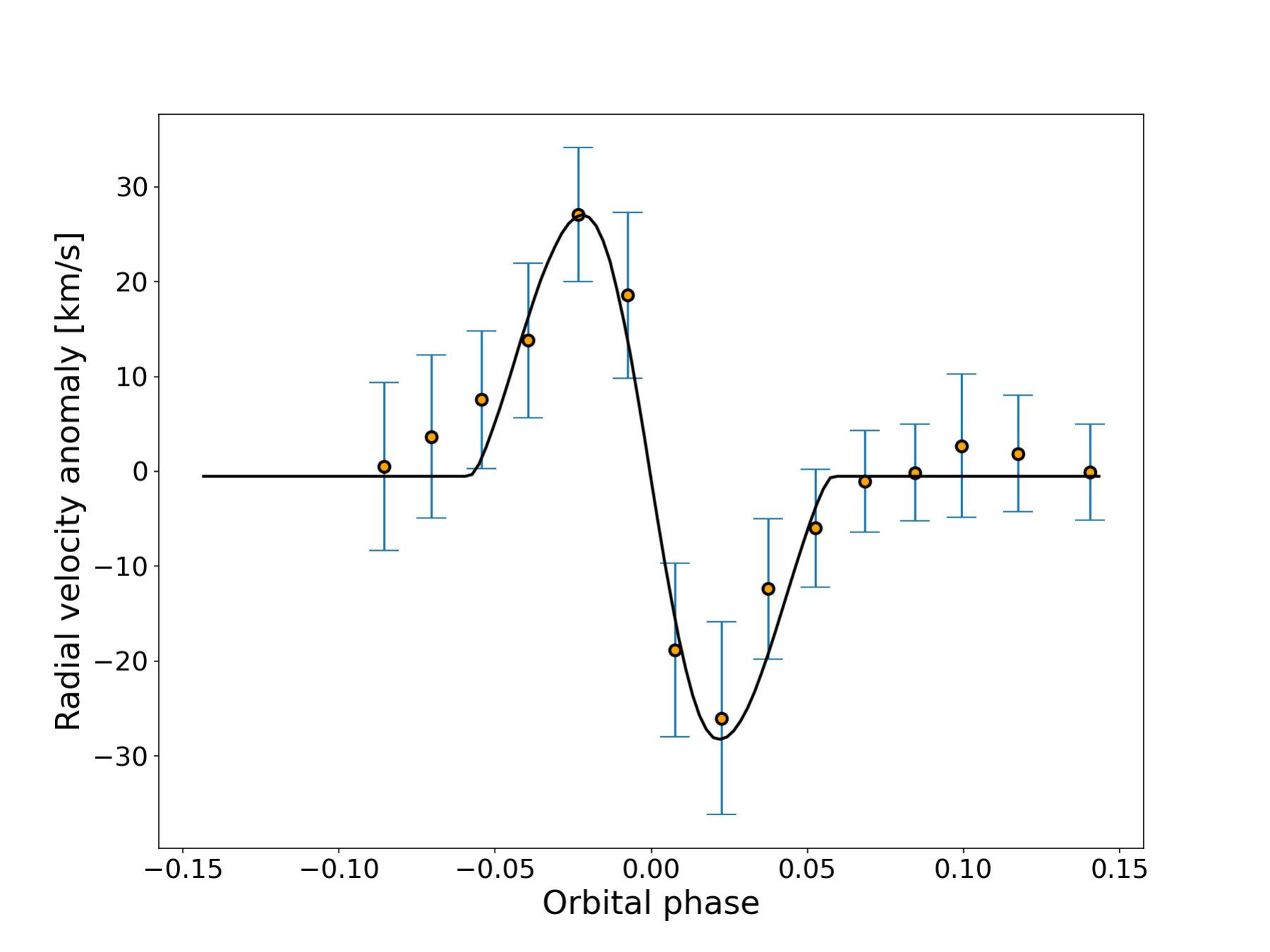}
    \caption{RV measurements with the RV orbital motion subtracted to show the RM signal (orange filled circles with blue error bars). Over-imposed (black line) is the fit of the RM effect. The data underlying the figure are available as Data behind the Figure.}
    \label{fig: fig1}
\end{figure}

We then subtracted the RV orbital signal from the original time series to isolate the RM signal. The RM effect was fitted with a purpose-made Python code with rotational velocity, relative size of the two stars, and the impact parameter as parameters. Given the large errors on the radial velocities, limb darkening was not considered, as its impact would not be visible. Due to the high symmetry of the curve, we concluded that the spin-orbit angle $\lambda\approx0$. This suggests that the strong tidal interactions expected from the very short orbital period have circularised and aligned the orbit, likely creating a tidally locked binary system. From this assumption, the stellar rotational period equals the orbital period, and the stars are in synchronous motion. This allowed us to infer a $v \sin i=183 \pm 3 \ \text{km s}^{-1}$. We obtained an error which is smaller than expected, considering the limitations of the S/N. This is likely due to an underestimation of the radius and period uncertainties, used to compute $v \sin i$. From the full-width half maximum (FWHM) of the lines, we derived a rotational broadening of approximately twice that estimated from synchronous motion -- possibly due to the blending of the lines of the two stars, which likely rotate with the same velocity. Therefore, we adopted the former value for fitting the RV data. We fit the RM function to the in-transit data points (see Fig. \ref{fig: fig1}) to derive the shape and amplitude of the curve.
From the RM semi-amplitude $\Delta V_{RM} =27\pm3 \ \text{km s}^{-1}$, the transit depth, and the $v \sin i$, we derived an impact parameter $b =0.55\pm 0.06$. 

Overall, our analysis supports a close, likely tidally locked binary system with a well-aligned orbit, consistent with strong tidal interactions shaping its dynamical configuration. However, the limited number of out-of-transit RV measurements introduces significant uncertainty in the companion mass estimate, highlighting the need for additional observations to robustly constrain the system’s fundamental parameters.  

\section{Searching for Barium enhancement in two blue straggler candidates of the M67 open cluster}


Blue straggler stars (BSSs) are known to occupy a specific region in color-magnitude diagrams (CMDs), appearing bluer or brighter than normal main-sequence (MS) turn-off stars of the cluster. A proposed formation pathway for a BSS is mass transfer (MT) from an evolved star onto a MS companion \cite[see, e.g.,][and refs. therein]{2015ASSL..413.....B}. In this scenario, the system is later observed as a long-period BSS with a white dwarf companion and an enrichment in s-process elements from the donor’s Asymptotic Giant Branch phase (e.g., \citealt{lugaro+12}). Detecting enhancement in elements such as Ba, Sr, Y in BSSs can thus provide important constraints on stellar evolution and the MT process. We report a spectroscopic analysis of the two long-period BSS candidates, NGC 2682 90 ($M_V = 10.82$; $P = 1221$ days, \citealt{nine+24}) and NGC 2682 124 ($M_V = 12.34$; $P=4913$~d, \citealt{nine+24}), of the open cluster M67 (4 Gyr old) to characterize their abundances and assess potential s-process enrichment.

\subsection{Dataset}

The observations were carried out on February 7th and 9th 2026 using the high-resolution ($R \sim 110,000$) HARPS spectrograph \citep{mayor+03} at the 3.6 m Telescope at La Silla Observatory. The individual spectra, covering from 3800 to 6900 \AA, were reduced with the official ESO HARPS pipeline (v. 3.3.12) -- which directly includes the barycentric Earth radial velocity correction -- and then co-added for each target, leading to an average signal-to-noise ratio per pixel of S/N $\simeq 70$ (total exposure time of 5400 s) and S/N $\simeq 16$ (1200 s) for NGC 2682 90 and NGC 2682 124, respectively.

\subsection{Analysis and Results}

After applying radial velocity corrections and a continuum normalization, we derived estimates for the atmospheric parameters of both targets using the {\tt iSpec} software \citep{ispec1, ispec2}, based on the fits of $\mathrm{Fe\,I}$ and $\mathrm{Fe\,II}$ lines (cross-checking with the H$\alpha$ and H$\beta$ line profiles). The synthetic template was generated with \texttt{SPECTRUM} \citep{spectrum}, adopting the model atmosphere \texttt{ATLAS} \citep{castelli+03} and solar abundances from \cite{grevesse+07}, and using the Gaia-ESO Survey v.6 line list \citep{ges_linelist}. We assumed $\rm{[M/H]}=0$ and treated the effective temperature $T_\mathrm{eff}$, the surface gravity $\log (g)$, the microturbulence $v_{\rm{mic}}$, and the projected rotational velocity $v \sin i$ as free parameters. 

For NGC 2682 90 we constrained $\log (g) = 3.42 \pm 0.21 $ and $v \sin i = 42.89 \pm 2.60$ km s$^{-1}$, compatible with results from \cite{nine+24}, although they adopted a higher microturbulence ($v_{\rm{mic}} = 2–2.5$ km s$^{-1}$) compared to our estimate of $v_{\rm{mic}} = 1.16 \pm 0.14$ km s$^{-1}$. Our estimated $T_{\rm{eff}} = 6566 \pm 38$ K agrees within 2$\sigma$ with their results. Then, we measured elemental abundances using spectral synthesis, obtaining: $A(\rm{Fe})=7.46 \pm 0.03$, $A(\rm{Mg})=7.53 \pm 0.05$, $A(\rm{Si})=7.35 \pm 0.12$, $A(\rm{Ca})=6.66 \pm 0.10$, $A(\rm{Ti})=4.79 \pm 0.16$, $A(\rm{Mn})=5.13 \pm 0.20$, all consistent within 2$\sigma$ with M67 MS stars \citep{souto+18}. No significant enrichment was reported from the Ba {\sc II} \ 5853.7 \AA\, line ([Ba/Fe]$=0.20 \pm 0.44$) -- however, the line was too weak to give a robust estimate. No Sr line was identified -- similarly, Y and Zr lines were also too weak for a reliable analysis.

\begin{figure*}[h!]
 \centering
\includegraphics[width=0.9\linewidth,angle=0]{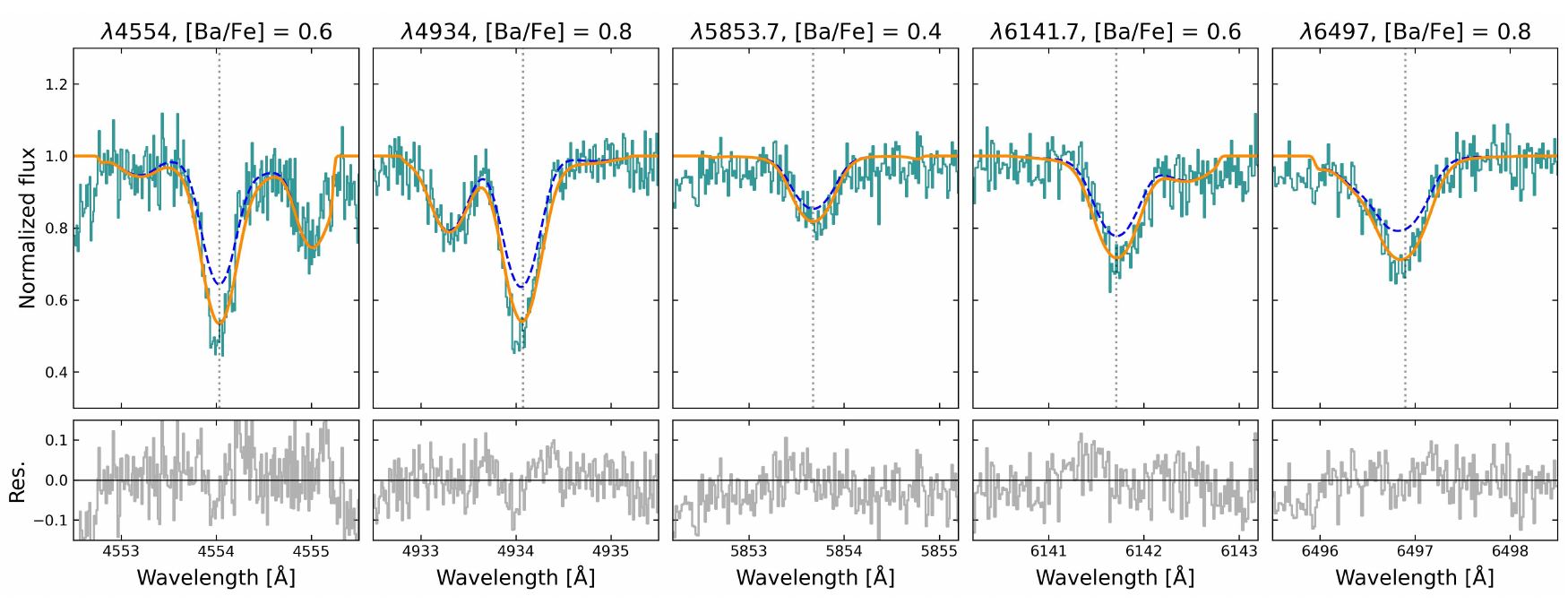}
\caption{$\mathrm{Ba\,II}$ lines in NGC 2682 124, assuming $v_{\rm{mic}} = 1.7$ km s$^{-1}$. The orange line is the best template matching the observed spectrum (green; residuals in grey), the blue line is the synthetic at [Ba/Fe]=0.}
\label{fig:general}
\end{figure*}

For NGC 2682 124, we constrained $T_{\rm{eff}} = 6513 \pm 80 $ K, $\log (g) = 3.78 \pm 0.22$, $v \sin i = 12.23 \pm 4.86$ km s$^{-1}$, all in good agreement with \cite{nine+24}, except for the microturbulence, for which they assume $v_{\rm{mic}} = 2–2.5$ km s$^{-1}$. We derived $v_{\rm{mic}} = 1.69 \pm 0.23$  km s$^{-1}$ -- reasonable for a star of $T_{\rm{eff}} \simeq$ 6500 K -- and obtained: $A(\rm{Fe}) = 7.48 \pm 0.03$, $A(\rm{C}) = 7.89 \pm 0.21$, $A(\rm{Mg}) = 7.55 \pm 0.08$, $A(\rm{Si}) = 7.29 \pm 0.07$, $A(\rm{Ca}) = 6.44 \pm 0.08$, $A(\rm{Ti}) = 5.32 \pm 0.09$, $A(\rm{Cr}) = 5.72 \pm 0.09$, and $A(\rm{Mn}) = 5.28 \pm 0.21$, generally consistent with \cite{souto+18}. Moreover, we performed an individual line synthesis with {\tt pymoog} \citep{pymoog} and focused on the available $\mathrm{Ba\,II}$ lines: 4554, 4934, 5853.7, 6141.7, 6497 \AA. We ran a grid of synthetic spectra varying [Ba/Fe] from $-0.2$ to $1.2$ with a step of 0.2 dex, for three values of $v_{\rm{mic}}$: 1.7, 2.0 and 2.5 km s$^{-1}$. 
For each $v_{\rm{mic}}$, the best-fit model minimized the reduced $\chi^2$, computed (assuming Poisson noise) as:

$$\chi^2 = \frac{1}{N} \sum_{i=1}^{N} \frac{(f_i - m_i)^2}{f_i},$$

\noindent where $f_i$ and $m_i$ are the observed and synthetic fluxes at pixel $i$, and $N$ is the number of pixels.

Therefore, we constrained a mean abundance of [Ba/Fe] = $0.64 \pm 0.07$, [Ba/Fe] = $0.48 \pm 0.09$, [Ba/Fe] = $0.16 \pm 0.07$ for $v_{\rm{mic}} = $ 1.7, 2.0 and 2.5 km s$^{-1}$, respectively. Interestingly, this result is in contradiction with \cite{nine+24}, who infer [Ba/Fe]$=-0.14$. This discrepancy may be due to the fact that they only use the weak $\mathrm{Ba\,II}$ 5853.7 \AA, whereas we also exploit stronger $\mathrm{Ba\,II}$ transitions from which it is possible to detect the deviation from solar abundance (Fig. \ref{fig:general}). Considering that [Fe/H] is approximately solar ([Fe/H]$\simeq 0$) for $v_{\rm{mic}}=$ 1.7 km s$^{-1}$, and decreases to [Fe/H]$=-0.10 \pm 0.04$ for $v_{\rm{mic}}=2.0$ km s$^{-1}$ and [Fe/H]$=-0.24 \pm 0.05$ for $v_{\rm{mic}}=2.5$ km s$^{-1}$, these results lead to a final average of [Ba/H]$=0.54 \pm 0.09$, indicating chemical enhancement. The spectrum also exhibits few lines of $\mathrm{Sr\,II}$, $\mathrm{Zr\,II}$ and $\mathrm{Y\,II}$, but no enrichment was determined by {\tt iSpec} ([Sr/Fe] = $0.28 \pm 0.20$, [Zr/Fe] = $0.10 \pm 0.47$, [Y/Fe] = $-0.11 \pm 0.32$).

Our analysis confirms that NGC 2682 124 exhibits significant Barium enrichment, likely linked to MT, while NGC 2682 90 shows no detectable s-process enhancement, highlighting diversity in the chemical signatures of long-period BSSs in M67.

\section{Spectroscopic detection of pulsations in the candidate $\delta$ Scuti star V845 Mon}

Blue stragglers (BS) are a class of rejuvenated stars easily identified in stellar clusters, formed through mass transfer, mergers, or stellar interactions \citep{McCrea,1976Hills}. They are located above and blueward of the main-sequence turnoff in the color–magnitude diagram \citep[CMD,][]{1953Sandage}. Many have been found in the classical instability strip, where $\delta$ Scuti or SX Phe oscillations are expected \citep{2007Arentoft}.

Rediet's star\footnote{‘Rediet’s star’ is an informal name we use for V845 Mon, in honour of Rediet Wauters, who passed away at a very young age.}
, V* V845 Mon, is located in the open cluster NGC 2506. It is a blue straggler and a candidate $\delta$ Scuti star with a $V_{mag}$= 14.510 \citep{2021Rain, gaia}. TESS photometric observations present a period of 0.0921 days, indicating strong short-period variability. Although BSs are often associated with binary systems, such a short period is unlikely to represent the orbital period, as an F-type primary would not fit within such a compact orbit. Instead, this periodicity is more consistent with $\delta$ Scuti-type pulsations. We present a time-resolved spectroscopic observations to confirm the $\delta$ Scuti classification and characterize its pulsation properties.

\subsection{Data} \label{sec:data}
The observations were carried out between February 7 and 9, 2026, using the ESO Faint Object Spectrograph and Camera (v.2) (EFOSC2) mounted on the New Technology Telescope (NTT) at the La Silla Observatory \citep{efosc2}.  Grism 19 and the 0.7$^{\prime \prime}$ slit were used, covering a wavelength range of 440--510 nm and providing a resolving power of $R \sim 2300$. An initial continuous monitoring session was performed, followed by two additional observations on later occasions, each with an exposure time of 180 s, resulting in a total of nine low-resolution spectra at median S/N of $\sim$\,12. The data were then reduced using the dedicated ESO pipeline run on the esorex tool \footnote{\url{https://www.eso.org/sci/software/cpl/esorex.html}}. The interstellar extinction was derived from $E(B-V) = 0.058 \pm 0.001$ \citep{AnthonyTwarog_2018}, resulting in $A_V = 0.178$. Considering the highly variable nature of the source, which may have affected its parallax measurement, the distance was derived using the cluster parallax, $\varpi = 0.292 \pm 0.002 \, \mathrm{mas}$ \citep{2020Cantat-Gaudin}. The absolute magnitude $M_{\rm V}$ was then calculated and compared with those reported in \cite{2013Pecaut}, to obtain the corresponding stellar parameters such as the effective temperature ($\mathrm{T_{eff} = 8600 K}$) and Bolometric Correction in V-band ($\mathrm{BC_V} = -0.04\,\mathrm{mag}$). Using these values, the stellar radius was then estimated to be $R = 1.827 \pm 0.016 \,R_\odot$.

\subsection{Analysis $\&$ Results} \label{sec:Analysis}

\begin{figure*}
    \centering
    \includegraphics[width=1\linewidth]{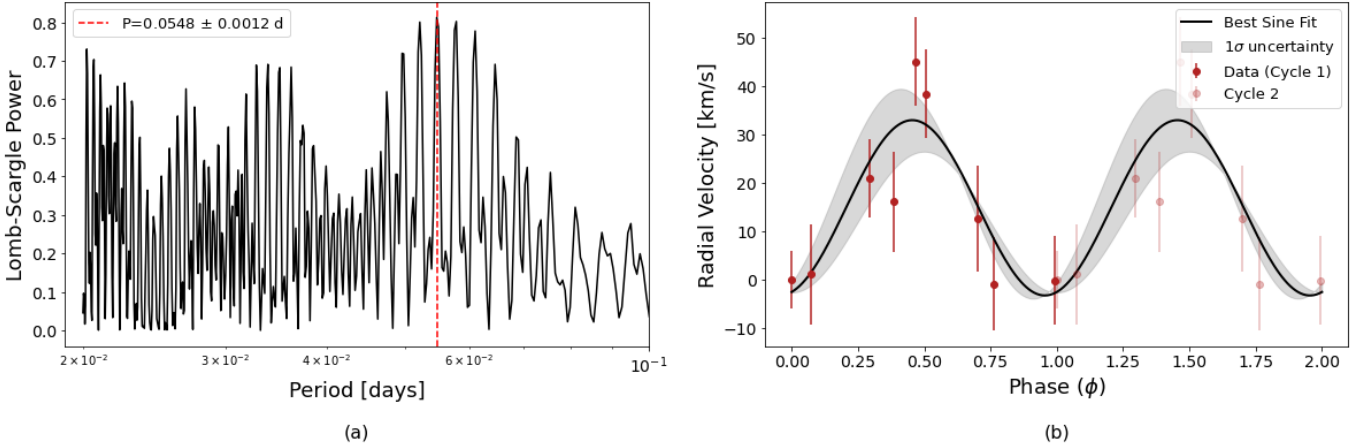}
    \caption{(a) Weighted Lomb–Scargle periodogram of the radial velocity measurements, showing a peak at $P = 0.0548$ days, marked by a red dashed line. (b) Phase-folded radial velocity curve using the derived period. The solid black line represents the best-fitting sinusoidal model, and the gray shaded region indicates the $1\sigma$ uncertainty of the fit. Error bars correspond to the estimated radial velocity uncertainties. \label{fig:figure}}
\end{figure*}

Taking into account the low spectral resolution, radial velocities were estimated based on the shift of the H$\beta$ line at 4861.33 \AA, which was the strongest feature for determining the variability in the spectra. A Lorentzian profile was fitted to the absorption line after defining a continuum-normalized wavelength window around the rest wavelength. The line centers were determined for each epoch, and relative radial velocities were estimated from the wavelength shifts with respect to the first observed spectrum, which was used as a template. Uncertainties were estimated assuming Poisson-based errors derived from the flux gradient across the line profile. A weighted Lomb–Scargle periodogram was computed over a period range of 0.02--2 days, giving a period of 0.0548 $\pm$ 0.0012 days. The uncertainty in the period was derived by propagating the frequency uncertainty estimated from the curvature of a parabola fitted to the highest peak of the periodogram. Using this period, a sinusoidal least-squares fit was applied to the phase-folded RV data, resulting in a semi-amplitude of 18.55 $\pm$ 3.65 $\mathrm{km\,s^{-1}}$ (Figure~\ref{fig:figure}). To estimate the systemic velocity, absolute radial velocities were calculated from the shift of the H$\beta$ line center relative to its vacuum wavelength, followed by Earth barycentric correction. The uncertainties were estimated from the covariance matrix of the fit and a systemic velocity of 77.15 $\pm$ 3.19 $\mathrm{km\,s^{-1}}$ was obtained. The cluster radial velocity of 84.29 $\pm$ 0.54 $\mathrm{km\,s^{-1}}$ \citep{Linck_2024} is consistent with this value within 2.2$\sigma$ uncertainties, providing support for its membership.

Interestingly, a mismatch is observed between the periods (P) derived from TESS photometry (0.0921 days) and spectroscopy (0.0548 $\pm$ 0.0012 days). To verify the nature of the observed variability, the period–luminosity relations from \cite{2011McNamara} and \cite{2024Poro}
were applied. The fundamental periods obtained were 0.0939 $\pm$ 0.0125 days and 0.0959 $\pm$ 0.0051 days, respectively, consistent with the TESS period, indicating that the luminosity variations are dominated by the fundamental mode. Using Equations 11, 12, and 13 from \cite{2024Poro}, the first, second, and third overtone periods were estimated to be 0.0752 $\pm$ 0.0068 days, 0.0491 $\pm$ 0.0151 days, and 0.0147 $\pm$ 0.0095 days, respectively. The period derived from the radial velocity analysis is consistent with the second overtone within 1$\sigma$. Furthermore, the ratio between the RV-derived period and the TESS period is $\sim$0.6, matching the expected 
$P_2/P_0$ ratio for second-overtone pulsators. This can be attributed to radial velocities tracing surface motions that are dominated by higher-frequency overtone modes.

Additionally, the ratio between the full radial velocity amplitude ($2K = 37.1$ km s$^{-1}$) and the TESS photometric amplitude ($\Delta m_V \approx 0.2$ mag) was estimated to be $\sim 185$ km s$^{-1}$ mag$^{-1}$, confirming the source as a strong radial pulsator consistent with a $\delta$ Scuti star \citep{1991Yang}. The radial displacement ($\Delta R$) was then estimated using the relation derived from \cite{2010Pedicelli}:
\begin{equation}
    \Delta R = \frac{p K P}{2\pi}
\end{equation}

Using the period-projection factor relation of \cite{2007Nardetto}, we estimate a projection factor  $p = 1.4568$,accounting for geometrical and dynamical effects in converting observed radial velocities into true pulsational velocities of the stellar surface. The resulting radial displacement corresponds to $1.6 \pm 0.32\%$ of the stellar radius, consistent with the range observed for high-amplitude $\delta$ Scuti stars (HADS). These results confirm the $\delta$ Scuti nature of this blue straggler star, which exhibits high-amplitude pulsations.


\section{Constraining the nature of the binary in the planetary nebula MPA J0705-1224}

Planetary nebulae (PNe) are generally thought to be the end stage of low-mass stars like our Sun. Recently, however, it became clear that this binarity plays an important role in the formation and evolution of PNe \citep{Boffin_2019}. One such binary in a PNe could be MPA J0705-1224 (PN G225.5-02.5), which has fist been proposed as a PN candidate by \citet{Miszalski_2008}. Recently, \citet{Chen_2025} determined the period of the variable binary in the center to be \SI{4.42}{\hour} using data from the Zwicky Transient Facility in g- and r-band filters.

\subsection{Observations \& Data Reduction}
We monitored the photometric variability of MPA J0705-1224 using EFOSC2, collecting 38 epochs of imaging data in the broadband Gunn-i filter (i\#705). Our observations span a baseline of 52 hours, with a central $~2.5$-hour sequence of 20-minute intervals. For validation of the binary and planetary nebulae, we collected a spectrum using GRISM\#11 with its second-order separation filter, GG375, and a \SI{1}{\arcsec} slit. The spectrum ranges from $338 - 752$ \si{\nm}, at a resolution of $\Delta \lambda = 15.8\,$\si{\angstrom}. Additionally, we collected standard calibration BIAS and SKY,FLAT images, as well as photometric (LTT2415) and spectro-photometric (SA95-107) standard stars. \\
For the photometric and spectroscopic data, we used the designated ESOReflex\footnote{\href{https://www.eso.org/sci/software/esoreflex/}{https://www.eso.org/sci/software/esoreflex/}} for the standard reductions and flux calibrations.
We extract the light curve using aperture photometry, making use of the \texttt{DAOStarFinder} function in \texttt{photutils}. We also extracted the flux of two reference targets (Gaia DR3 3045709468789022464 \& Gaia DR2 3044958742866363264), which show no variability, in order to perform relative photometry. For the spectral extraction, we started with the 2D spectrum output from ESOReflex, allowing us to separate the nebula from the central binary star. We fit a Gaussian to the spectral trace profile, using only the central rows for the stellar spectrum and only the columns after the Gaussian contribution falls below 1\% of its maximum for the nebula. Lastly, we clean the binary spectrum by subtracting the nebula spectrum.  

\subsection{Results}
\begin{figure*}[ht!]
\centering
    \makebox[\textwidth][c]{
    \includegraphics[width=\textwidth]{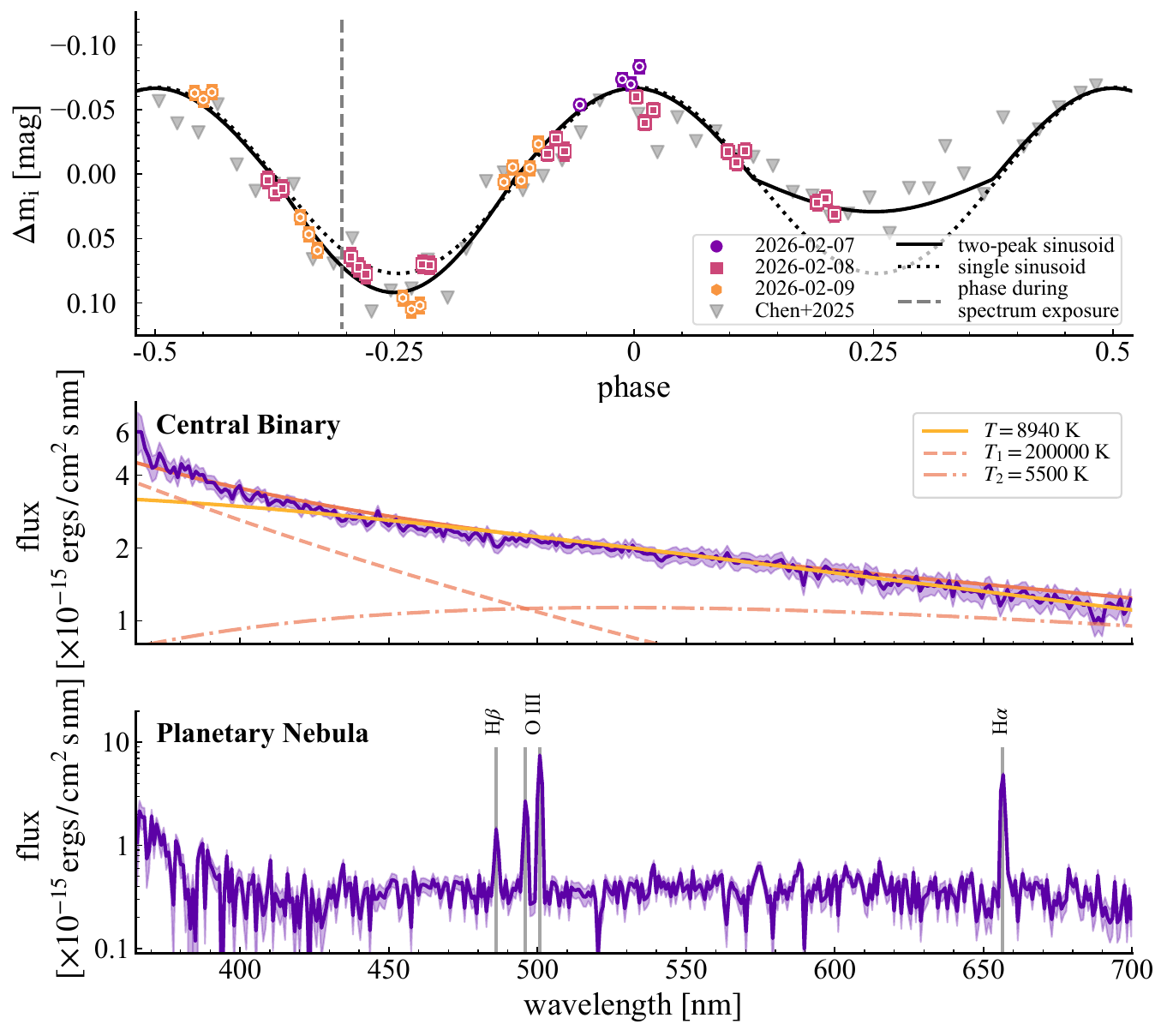}}
\caption{top: Phase folded light curve of MPA J0705-1224. Observation nights are indicated by different markers and colors; the vertical dashed line marks the phase during the spectrum shown in the panels below. middle: Spectrum of the binary (purple), fit with a single (yellow) and a double (orange) blackbody model, the two individual blackbody contributions to the double blackbody model are shown by the dashed and dashed-dotted lines. bottom: Planetary nebula spectrum (purple), H$\mathrm{\alpha}$, H$\mathrm{\beta}$ and the OIII doublet emission lines are indicated.
\label{fig: spectra}}
\end{figure*}

The differential photometry of MPA J0705-1224 shows regular variability. A periodogram shows two peaks at $\approx 2.21$\,\si{\hour} and $\approx 2.45\,$\si{\hour}, of which the latter is attributed to the length of our central sequence and is therefore neglected for further analysis. To assess the peak at $\approx 2.21\,$\si{\hour}, we fit a sinusoidal light curve, achieving a $\chi^2_\text{red} = 11.6$ for a period of P = \SI{2.207(0.003)}{\hour}, hinting imperfections of the model. This is also evident when comparing the model to the data points (see \autoref{fig: spectra} top), where a clear change in depth between the negative lobes is visible. We therefore also adopt a double-sine fit, allowing a single positive and two distinct negative amplitudes as three free parameters. This fit, with now double the initial period of P = \SI{4.42(0.08)}{\hour}, yields a $\chi^2_\text{red} = 5.0$, which is preferred over the single sinusoidal curve but still not fully accurate. Our measured photometry also agrees with the recently published ZTF data by \citet{Chen_2025}.\\
\newline
The nebula spectrum, \autoref{fig: spectra} (bottom), shows a clear H$\mathrm{\alpha}$ line as well as the OIII doublet emission, but only a faint signal from H$\mathrm{\beta}$. Fitting Gaussian profiles to the H$\mathrm{\beta}$ and H$\mathrm{\alpha}$ emission lines, we derive the integrated line fluxes to:
\begin{align*}
    \mathrm{I}_\mathrm{H\beta} &= \SI{17(4)}{\times 10^{-16} \, \ergs \, \s^{-1} \cm^{-2}}\\
    \mathrm{I}_\mathrm{H\alpha} &= \SI{71(11)}{\times 10^{-16} \ergs \, \s^{-1} \cm^{-2}},
\end{align*}
respectively. Following \citet{Wesson_2004} we compute the visual extinction to $\mathrm{A}_\mathrm{v} = (1.2 \pm 0.8)$ \si{mag}. Further, assuming a galactic extinction law with $R_V = 3.1$ \citep{Cardelli_1989}, we derive $\text{E(B-V)} = (0.39\pm0.26)$ \si{mag}.\\
\newline
The stellar spectrum, \autoref{fig: spectra} (center), deviates from a single blackbody. Looking at the individual spectral rows before integration, it becomes clear that the central spectrum is composed of two distinct spectral shapes. We therefore fit the central stars with two individual blackbody functions, allowing the temperatures and intensities to vary. The double black body model ($\mathrm{T}_1 \approx 200\,000 \, \text{K}, \mathrm{T}_2 \approx 5500 \,\text{K}$) is favored with a $\chi^2_\text{red} = 1.07$ over the single blackbody ($\mathrm{T} \approx 8940 \, \text{K}$) with a $\chi^2_\text{red} = 2.50$.

\subsection{Conclusion}
We can confirm the nature of MPA J0705-1224 as a true PN based on spectral line identification due to the detection of the OIII doublet as well as H$\alpha$ and H$\beta$ emission lines. Using the intensity ratio ratio 
$\mathrm{I}_\mathrm{H\alpha}/\mathrm{I}_\mathrm{H\beta} = 4.3 \pm 1.1$ we compute $\mathrm{A}_\mathrm{v} = (1.2 \pm 0.8)$ \si{mag} and $\text{E(B-V)} = (0.39\pm0.26)$ \si{mag}, assuming $R_V = 3.1$ \citep{Cardelli_1989}. The large error reflects the lower S/N in the blue part of the spectrum, which could systematically affect the measurement of the H$\beta$ line.\\
\newline
From our retrieved period of P = \SI{4.42(0.08)}{\hour} and the small change in apparent brightness, together with two distinct blackbody temperatures, we conclude that in the case of MPA J0705-1224, the central binary is composed of a white dwarf and a secondary star, which is tidally elongated by the gravitational force of the white dwarf. The orbital period of the binary implies a maximum Roche lobe radius of the companion of $0.55 R_\odot$, and thus would only be able to accommodate a $0.5 - 0.6\, M_\odot$ star corresponding to an M0V type star of $T \approx 3500K$. Hence, the measured temperature of $T \approx 5500 K$ can only be achieved if the secondary star is heated by the irradiation of the primary white dwarf. However, assuming that the difference in surface temperature is then causing the variability would imply a brightness difference of at least 1 mag. As we do not observe such a change in brightness, resolving this quite puzzling result requires further observations.

\begin{acknowledgments}\label{Acknowledgment}
This project was undertaken during the ESO La Silla Observing School 2026. We are extremely grateful to the organisers of the School, as well as to the staff of the La Silla Observatory for their technical support. SDS thanks Federico Rizzuti and Linda Lombardo for helpful discussions. 
This work has made use of data from the European Space Agency (ESA) mission Gaia, processed by the Gaia Data Processing and Analysis Consortium (DPAC). This work makes use of data from the {\it TESS} mission, obtained from the Data Release 2 (DR2; 2020-05-27) of the PATHOS repository \citep{10.17909/t9-es7m-vw14} in the Mikulski Archive for Space Telescopes (MAST).
\end{acknowledgments}

\facilities{ESO NTT 3.5m (EFOSC2), ESO 3.6m (HARPS, NIRPS)}

\software{astropy \citep{AstropyCollaboration_2022}, 
          photutils \citep{Bradley_2025}
          numpy \cite{Harris_2020},
          scipy \citep{2020SciPy-NMeth}
          matplotlib \citep{Hunter:2007},
          esorex \& esoreflex \citep{ESOCPLDevelopmentTeam_2015},   
          }

\bibliography{newbibliography}{}

@INPROCEEDINGS{castelli+03,
       author = {{Castelli}, F. and {Kurucz}, R.~L.},
        title = "{New Grids of ATLAS9 Model Atmospheres}",
    booktitle = {Modelling of Stellar Atmospheres},
         year = 2003,
       editor = {{Piskunov}, N. and {Weiss}, W.~W. and {Gray}, D.~F.},
       series = {IAU Symposium},
       volume = {210},
        month = jan,
        pages = {A20},
          doi = {10.48550/arXiv.astro-ph/0405087},
archivePrefix = {arXiv},
       eprint = {astro-ph/0405087},
 primaryClass = {astro-ph},
       adsurl = {https://ui.adsabs.harvard.edu/abs/2003IAUS..210P.A20C}
}

@PROCEEDINGS{2015ASSL..413.....B,
        title = "{Ecology of Blue Straggler Stars}",
    booktitle = {Astrophysics and Space Science Library},
         year = 2015,
       editor = {{Boffin}, Henri M.~J. and {Carraro}, Giovanni and {Beccari}, Giacomo},
       series = {Astrophysics and Space Science Library},
       volume = {413},
        month = jan,
          doi = {10.1007/978-3-662-44434-4},
archivePrefix = {arXiv},
       eprint = {1406.3909},
 primaryClass = {astro-ph.SR},
       adsurl = {https://ui.adsabs.harvard.edu/abs/2015ASSL..413.....B}
}

@ARTICLE{ges_linelist,
       author = {{Heiter}, U. and {Lind}, K. and {Asplund}, M. and {Barklem}, P.~S. and {Bergemann}, M. and {Magrini}, L. and {Masseron}, T. and {Mikolaitis}, {\v{S}}. and {Pickering}, J.~C. and {Ruffoni}, M.~P.},
        title = "{Atomic and molecular data for optical stellar spectroscopy}",
      journal = {\physscr},
         year = 2015,
        month = may,
       volume = {90},
       number = {5},
          eid = {054010},
        pages = {054010},
          doi = {10.1088/0031-8949/90/5/054010},
archivePrefix = {arXiv},
       eprint = {1506.06697},
 primaryClass = {astro-ph.IM},
       adsurl = {https://ui.adsabs.harvard.edu/abs/2015PhyS...90e4010H}
}

@ARTICLE{grevesse+07,
       author = {{Grevesse}, N. and {Asplund}, M. and {Sauval}, A.~J.},
        title = "{The Solar Chemical Composition}",
      journal = {\ssr},
         year = 2007,
        month = jun,
       volume = {130},
       number = {1-4},
        pages = {105-114},
          doi = {10.1007/s11214-007-9173-7},
       adsurl = {https://ui.adsabs.harvard.edu/abs/2007SSRv..130..105G}
}

@ARTICLE{ispec1,
       author = {{Blanco-Cuaresma}, S. and {Soubiran}, C. and {Heiter}, U. and {Jofr{\'e}}, P.},
        title = "{Determining stellar atmospheric parameters and chemical abundances of FGK stars with iSpec}",
      journal = {\aap},
         year = 2014,
        month = sep,
       volume = {569},
          eid = {A111},
        pages = {A111},
          doi = {10.1051/0004-6361/201423945},
archivePrefix = {arXiv},
       eprint = {1407.2608},
 primaryClass = {astro-ph.IM},
       adsurl = {https://ui.adsabs.harvard.edu/abs/2014A&A...569A.111B}
}

@ARTICLE{ispec2,
       author = {{Blanco-Cuaresma}, Sergi},
        title = "{Modern stellar spectroscopy caveats}",
      journal = {\mnras},
         year = 2019,
        month = jun,
       volume = {486},
       number = {2},
        pages = {2075-2101},
          doi = {10.1093/mnras/stz549},
archivePrefix = {arXiv},
       eprint = {1902.09558},
 primaryClass = {astro-ph.SR},
       adsurl = {https://ui.adsabs.harvard.edu/abs/2019MNRAS.486.2075B}
}

@ARTICLE{lugaro+12,
       author = {{Lugaro}, Maria and {Karakas}, Amanda I. and {Stancliffe}, Richard J. and {Rijs}, Carlos},
        title = "{The s-process in Asymptotic Giant Branch Stars of Low Metallicity and the Composition of Carbon-enhanced Metal-poor Stars}",
      journal = {\apj},
         year = 2012,
        month = mar,
       volume = {747},
       number = {1},
          eid = {2},
        pages = {2},
          doi = {10.1088/0004-637X/747/1/2},
archivePrefix = {arXiv},
       eprint = {1112.2757},
 primaryClass = {astro-ph.SR},
       adsurl = {https://ui.adsabs.harvard.edu/abs/2012ApJ...747....2L}
}

@ARTICLE{mayor+03,
       author = {{Mayor}, M. and {Pepe}, F. and {Queloz}, D. and {Bouchy}, F. and {Rupprecht}, G. and {Lo Curto}, G. and {Avila}, G. and {Benz}, W. and {Bertaux}, J.-L. and {Bonfils}, X. and {Dall}, Th. and {Dekker}, H. and {Delabre}, B. and {Eckert}, W. and {Fleury}, M. and {Gilliotte}, A. and {Gojak}, D. and {Guzman}, J.~C. and {Kohler}, D. and {Lizon}, J.-L. and {Longinotti}, A. and {Lovis}, C. and {Megevand}, D. and {Pasquini}, L. and {Reyes}, J. and {Sivan}, J.-P. and {Sosnowska}, D. and {Soto}, R. and {Udry}, S. and {van Kesteren}, A. and {Weber}, L. and {Weilenmann}, U.},
        title = "{Setting New Standards with HARPS}",
      journal = {The Messenger},
         year = 2003,
        month = dec,
       volume = {114},
        pages = {20-24},
       adsurl = {https://ui.adsabs.harvard.edu/abs/2003Msngr.114...20M}
}

@ARTICLE{nine+24,
       author = {{Nine}, Andrew C. and {Mathieu}, Robert D. and {Schuler}, Simon C. and {Milliman}, Katelyn E.},
        title = "{WIYN Open Cluster Study. XC. Barium Surface Abundances of Blue Straggler Stars in the Open Clusters NGC 7789 and M67}",
      journal = {\apj},
         year = 2024,
        month = aug,
       volume = {970},
       number = {2},
          eid = {187},
        pages = {187},
          doi = {10.3847/1538-4357/ad534b},
archivePrefix = {arXiv},
       eprint = {2405.20242},
 primaryClass = {astro-ph.SR},
       adsurl = {https://ui.adsabs.harvard.edu/abs/2024ApJ...970..187N}
}

@INPROCEEDINGS{pymoog,
       author = {{Jian}, Mingjie},
        title = "{pymoog}",
    booktitle = {Zenodo Software},
         year = 2024,
       volume = {75},
        month = apr,
    publisher = {Zenodo},
          eid = {7582434},
        pages = {7582434},
          doi = {10.5281/zenodo.7582434},
       adsurl = {https://ui.adsabs.harvard.edu/abs/2024zndo...7582434J}
}

@ARTICLE{souto+18,
       author = {{Souto}, Diogo and {Cunha}, Katia and {Smith}, Verne V. and {Allende Prieto}, C. and {Garc{\'\i}a-Hern{\'a}ndez}, D.~A. and {Pinsonneault}, Marc and {Holzer}, Parker and {Frinchaboy}, Peter and {Holtzman}, Jon and {Johnson}, J.~A. and {J{\"o}nsson}, Henrik and {Majewski}, Steven R. and {Shetrone}, Matthew and {Sobeck}, Jennifer and {Stringfellow}, Guy and {Teske}, Johanna and {Zamora}, Olga and {Zasowski}, Gail and {Carrera}, Ricardo and {Stassun}, Keivan and {Fernandez-Trincado}, J.~G. and {Villanova}, Sandro and {Minniti}, Dante and {Santana}, Felipe},
        title = "{Chemical Abundances of Main-sequence, Turnoff, Subgiant, and Red Giant Stars from APOGEE Spectra. I. Signatures of Diffusion in the Open Cluster M67}",
      journal = {\apj},
         year = 2018,
        month = apr,
       volume = {857},
       number = {1},
          eid = {14},
        pages = {14},
          doi = {10.3847/1538-4357/aab612},
archivePrefix = {arXiv},
       eprint = {1803.04461},
 primaryClass = {astro-ph.SR},
       adsurl = {https://ui.adsabs.harvard.edu/abs/2018ApJ...857...14S}
}

@software{spectrum,
       author = {{Gray}, Richard O.},
        title = "{SPECTRUM: A stellar spectral synthesis program}",
 howpublished = {Astrophysics Source Code Library, record ascl:9910.002},
         year = 1999,
        month = oct,
          eid = {ascl:9910.002},
archivePrefix = {ascl},
       eprint = {9910.002},
       adsurl = {https://ui.adsabs.harvard.edu/abs/1999ascl.soft10002G}
}

@ARTICLE{Rossiter1924,
       author = {{Rossiter}, R.~A.},
        title = "{On the detection of an effect of rotation during eclipse in the velocity of the brighter component of beta Lyrae, and on the constancy of velocity of this system.}",
      journal = {\apj},
         year = 1924,
        month = jul,
       volume = {60},
        pages = {15-21},
          doi = {10.1086/142825},
       adsurl = {https://ui.adsabs.harvard.edu/abs/1924ApJ....60...15R}
}

@ARTICLE{2025AJ....170..274Z,
       author = {{Zak}, J. and {Boffin}, H.~M.~J. and {Bocchieri}, A. and {Sedaghati}, E. and {Balkoova}, Z. and {Kabath}, P.},
        title = "{Ten Aligned Orbits: Planet Migration in the Era of JWST and Ariel}",
      journal = {\aj},
         year = 2025,
        month = nov,
       volume = {170},
       number = {5},
          eid = {274},
        pages = {274},
          doi = {10.3847/1538-3881/ae071b},
archivePrefix = {arXiv},
       eprint = {2505.20516},
 primaryClass = {astro-ph.EP},
       adsurl = {https://ui.adsabs.harvard.edu/abs/2025AJ....170..274Z}
}

@ARTICLE{McLaughlin1924,
       author = {{McLaughlin}, D.~B.},
        title = "{Some results of a spectrographic study of the Algol system.}",
      journal = {\apj},
         year = 1924,
        month = jul,
       volume = {60},
        pages = {22-31},
          doi = {10.1086/142826},
       adsurl = {https://ui.adsabs.harvard.edu/abs/1924ApJ....60...22M}
}

@ARTICLE{Queloz2000,
       author = {{Queloz}, D. and {Eggenberger}, A. and {Mayor}, M. and {Perrier}, C. and {Beuzit}, J.~L. and {Naef}, D. and {Sivan}, J.~P. and {Udry}, S.},
        title = "{Detection of a spectroscopic transit by the planet orbiting the star HD209458}",
      journal = {\aap},
         year = 2000,
        month = jul,
       volume = {359},
        pages = {L13-L17},
          doi = {10.48550/arXiv.astro-ph/0006213},
archivePrefix = {arXiv},
       eprint = {astro-ph/0006213},
 primaryClass = {astro-ph},
       adsurl = {https://ui.adsabs.harvard.edu/abs/2000A&A...359L..13Q}
}

@ARTICLE{Johnson2009,
       author = {{Johnson}, John Asher and {Winn}, Joshua N. and {Albrecht}, Simon and {Howard}, Andrew W. and {Marcy}, Geoffrey W. and {Gazak}, J. Zachary},
        title = "{A Third Exoplanetary System with Misaligned Orbital and Stellar Spin Axes}",
      journal = {\pasp},
         year = 2009,
        month = oct,
       volume = {121},
       number = {884},
        pages = {1104},
          doi = {10.1086/644604},
archivePrefix = {arXiv},
       eprint = {0907.5204},
 primaryClass = {astro-ph.EP},
       adsurl = {https://ui.adsabs.harvard.edu/abs/2009PASP..121.1104J}
}

@ARTICLE{Lendl2014,
       author = {{Lendl}, M. and {Triaud}, A.~H.~M.~J. and {Anderson}, D.~R. and {Collier Cameron}, A. and {Delrez}, L. and {Doyle}, A.~P. and {Gillon}, M. and {Hellier}, C. and {Jehin}, E. and {Maxted}, P.~F.~L. and {Neveu-VanMalle}, M. and {Pepe}, F. and {Pollacco}, D. and {Queloz}, D. and {S{\'e}gransan}, D. and {Smalley}, B. and {Smith}, A.~M.~S. and {Udry}, S. and {Van Grootel}, V. and {West}, R.~G.},
        title = "{WASP-117b: a 10-day-period Saturn in an eccentric and misaligned orbit}",
      journal = {\aap},
         year = 2014,
        month = aug,
       volume = {568},
          eid = {A81},
        pages = {A81},
          doi = {10.1051/0004-6361/201424481},
archivePrefix = {arXiv},
       eprint = {1406.6942},
 primaryClass = {astro-ph.EP},
       adsurl = {https://ui.adsabs.harvard.edu/abs/2014A&A...568A..81L}
}

@ARTICLE{Winn2010,
       author = {{Winn}, Joshua N. and {Fabrycky}, Daniel and {Albrecht}, Simon and {Johnson}, John Asher},
        title = "{Hot Stars with Hot Jupiters Have High Obliquities}",
      journal = {\apjl},
         year = 2010,
        month = aug,
       volume = {718},
       number = {2},
        pages = {L145-L149},
          doi = {10.1088/2041-8205/718/2/L145},
archivePrefix = {arXiv},
       eprint = {1006.4161},
 primaryClass = {astro-ph.EP},
       adsurl = {https://ui.adsabs.harvard.edu/abs/2010ApJ...718L.145W}
}

@ARTICLE{Triaud2011,
       author = {{Triaud}, A.~H.~M.~J.},
        title = "{The time dependence of hot Jupiters' orbital inclinations}",
      journal = {\aap},
         year = 2011,
        month = oct,
       volume = {534},
          eid = {L6},
        pages = {L6},
          doi = {10.1051/0004-6361/201117713},
archivePrefix = {arXiv},
       eprint = {1109.5813},
 primaryClass = {astro-ph.EP},
       adsurl = {https://ui.adsabs.harvard.edu/abs/2011A&A...534L...6T}
}

@ARTICLE{Albrecht2012,
       author = {{Albrecht}, Simon and {Winn}, Joshua N. and {Johnson}, John A. and {Howard}, Andrew W. and {Marcy}, Geoffrey W. and {Butler}, R. Paul and {Arriagada}, Pamela and {Crane}, Jeffrey D. and {Shectman}, Stephen A. and {Thompson}, Ian B. and {Hirano}, Teruyuki and {Bakos}, Gaspar and {Hartman}, Joel D.},
        title = "{Obliquities of Hot Jupiter Host Stars: Evidence for Tidal Interactions and Primordial Misalignments}",
      journal = {\apj},
         year = 2012,
        month = sep,
       volume = {757},
       number = {1},
          eid = {18},
        pages = {18},
          doi = {10.1088/0004-637X/757/1/18},
archivePrefix = {arXiv},
       eprint = {1206.6105},
 primaryClass = {astro-ph.SR},
       adsurl = {https://ui.adsabs.harvard.edu/abs/2012ApJ...757...18A}
}

@INPROCEEDINGS{Triaud2014,
       author = {{Triaud}, A.~H.~M.~J. and {Collier Cameron}, A. and {Queloz}, D. and {Anderson}, D.~R. and {Brown}, D.~J.~A. and {Smalley}, B. and {Bouchy}, F. and {Lendl}, M. and {Gillon}, M.},
        title = "{Spin-Orbit Angles as a Probe to Orbital Evolution}",
    booktitle = {Exploring the Formation and Evolution of Planetary Systems},
         year = 2014,
       editor = {{Booth}, Mark and {Matthews}, Brenda C. and {Graham}, James R.},
       series = {IAU Symposium},
       volume = {299},
        month = jan,
        pages = {399-400},
          doi = {10.1017/S174392131300759X},
       adsurl = {https://ui.adsabs.harvard.edu/abs/2014IAUS..299..399T}
}

@ARTICLE{Triaud2017,
       author = {{Triaud}, Amaury H.~M.~J. and {Martin}, David V. and {S{\'e}gransan}, Damien and {Smalley}, Barry and {Maxted}, Pierre F.~L. and {Anderson}, David R. and {Bouchy}, Fran{\c{c}}ois and {Collier Cameron}, Andrew and {Faedi}, Francesca and {G{\'o}mez Maqueo Chew}, Yilen and {Hebb}, Leslie and {Hellier}, Coel and {Marmier}, Maxime and {Pepe}, Francesco and {Pollacco}, Don and {Queloz}, Didier and {Udry}, St{\'e}phane and {West}, Richard},
        title = "{The EBLM Project. IV. Spectroscopic orbits of over 100 eclipsing M dwarfs masquerading as transiting hot Jupiters}",
      journal = {\aap},
         year = 2017,
        month = dec,
       volume = {608},
          eid = {A129},
        pages = {A129},
          doi = {10.1051/0004-6361/201730993},
archivePrefix = {arXiv},
       eprint = {1707.07521},
 primaryClass = {astro-ph.SR},
       adsurl = {https://ui.adsabs.harvard.edu/abs/2017A&A...608A.129T}
}

@ARTICLE{Mayor2003,
       author = {{Mayor}, M. and {Pepe}, F. and {Queloz}, D. and {Bouchy}, F. and {Rupprecht}, G. and {Lo Curto}, G. and {Avila}, G. and {Benz}, W. and {Bertaux}, J.-L. and {Bonfils}, X. and {Dall}, Th. and {Dekker}, H. and {Delabre}, B. and {Eckert}, W. and {Fleury}, M. and {Gilliotte}, A. and {Gojak}, D. and {Guzman}, J.~C. and {Kohler}, D. and {Lizon}, J.-L. and {Longinotti}, A. and {Lovis}, C. and {Megevand}, D. and {Pasquini}, L. and {Reyes}, J. and {Sivan}, J.-P. and {Sosnowska}, D. and {Soto}, R. and {Udry}, S. and {van Kesteren}, A. and {Weber}, L. and {Weilenmann}, U.},
        title = "{Setting New Standards with HARPS}",
      journal = {The Messenger},
         year = 2003,
        month = dec,
       volume = {114},
        pages = {20-24},
       adsurl = {https://ui.adsabs.harvard.edu/abs/2003Msngr.114...20M}
}

@ARTICLE{Miszalski_2008,
       author = {{Miszalski}, Brent and {Parker}, Q.~A. and {Acker}, A. and {Birkby}, J.~L. and {Frew}, D.~J. and {Kovacevic}, A.},
        title = "{MASH-II: more planetary nebulae from the AAO/UKST H{\ensuremath{\alpha}} survey}",
      journal = {\mnras},
         year = 2008,
       volume = {384},
       number = {2},
        pages = {525-534},
          doi = {10.1111/j.1365-2966.2007.12727.x},
}

@ARTICLE{Chen_2025,
       author = {{Chen}, Pinjian and {Fang}, Xuan and {Chen}, Xiaodian and {Liu}, Jifeng},
        title = "{Periodic Variability of the Central Stars of Planetary Nebulae Surveyed through the Zwicky Transient Facility}",
      journal = {\apj},
         year = 2025,
        month = feb,
       volume = {980},
       number = {2},
          eid = {227},
        pages = {227},
          doi = {10.3847/1538-4357/ada94a},
archivePrefix = {arXiv},
       eprint = {2501.06056},
 primaryClass = {astro-ph.SR},
       adsurl = {https://ui.adsabs.harvard.edu/abs/2025ApJ...980..227C}
}

@ARTICLE{Cardelli_1989,
       author = {{Cardelli}, Jason A. and {Clayton}, Geoffrey C. and {Mathis}, John S.},
        title = "{The Relationship between Infrared, Optical, and Ultraviolet Extinction}",
      journal = {\apj},
         year = 1989,
        month = oct,
       volume = {345},
        pages = {245},
          doi = {10.1086/167900},
       adsurl = {https://ui.adsabs.harvard.edu/abs/1989ApJ...345..245C}
}

@BOOK{Boffin_2019,
       author = {{Boffin}, Henri M.~J. and {Jones}, David},
        title = "{The Importance of Binaries in the Formation and Evolution of Planetary Nebulae}",
         year = 2019,
          doi = {10.1007/978-3-030-25059-1},
       adsurl = {https://ui.adsabs.harvard.edu/abs/2019ibfe.book.....B}
}

@article{Bradley_2025,
       author = {{Bradley}, Larry and {Sip{\H{o}}cz}, Brigitta and {Robitaille}, Thomas and {Tollerud}, Erik and {Vin{\'\i}cius}, Z{\'e} and {Deil}, Christoph and {Barbary}, Kyle and {Wilson}, Tom J and {Busko}, Ivo and {Donath}, Axel and {G{\"u}nther}, Hans Moritz and {Cara}, Mihai and {Lim}, P.~L. and {Me{\ss}linger}, Sebastian and {Burnett}, Zach and {Conseil}, Simon and {Droettboom}, Michael and {Bostroem}, Azalee and {Bray}, E.~M. and {Andersen Bratholm}, Lars and {Jamieson}, William and {Ginsburg}, Adam and {Barentsen}, Geert and {Craig}, Matt and {Pascual}, Sergio and {Rathi}, Shivangee and {Perrin}, Marshall and {Morris}, Brett M.},
        title = "{astropy/photutils: 2.2.0}",
         year = 2025,
        month = feb,
          eid = {10.5281/zenodo.14889440},
          doi = {10.5281/zenodo.14889440},
      version = {2.2.0},
    publisher = {Zenodo},
       adsurl = {https://ui.adsabs.harvard.edu/abs/2025zndo..14889440B}
}

@ARTICLE{AstropyCollaboration_2022,
       author = {{Astropy Collaboration} and {Price-Whelan}, Adrian M. and {Lim}, Pey Lian and {Earl}, Nicholas and {Starkman}, Nathaniel and {Bradley}, Larry and {Shupe}, David L. and {Patil}, Aarya A. and {Corrales}, Lia and {Brasseur}, C.~E. and {N{\"o}the}, Maximilian and {Donath}, Axel and {Tollerud}, Erik and {Morris}, Brett M. and {Ginsburg}, Adam and {Vaher}, Eero and {Weaver}, Benjamin A. and {Tocknell}, James and {Jamieson}, William and {van Kerkwijk}, Marten H. and {Robitaille}, Thomas P. and {Merry}, Bruce and {Bachetti}, Matteo and {G{\"u}nther}, H. Moritz and {Aldcroft}, Thomas L. and {Alvarado-Montes}, Jaime A. and {Archibald}, Anne M. and {B{\'o}di}, Attila and {Bapat}, Shreyas and {Barentsen}, Geert and {Baz{\'a}n}, Juanjo and {Biswas}, Manish and {Boquien}, M{\'e}d{\'e}ric and {Burke}, D.~J. and {Cara}, Daria and {Cara}, Mihai and {Conroy}, Kyle E. and {Conseil}, Simon and {Craig}, Matthew W. and {Cross}, Robert M. and {Cruz}, Kelle L. and {D'Eugenio}, Francesco and {Dencheva}, Nadia and {Devillepoix}, Hadrien A.~R. and {Dietrich}, J{\"o}rg P. and {Eigenbrot}, Arthur Davis and {Erben}, Thomas and {Ferreira}, Leonardo and {Foreman-Mackey}, Daniel and {Fox}, Ryan and {Freij}, Nabil and {Garg}, Suyog and {Geda}, Robel and {Glattly}, Lauren and {Gondhalekar}, Yash and {Gordon}, Karl D. and {Grant}, David and {Greenfield}, Perry and {Groener}, Austen M. and {Guest}, Steve and {Gurovich}, Sebastian and {Handberg}, Rasmus and {Hart}, Akeem and {Hatfield-Dodds}, Zac and {Homeier}, Derek and {Hosseinzadeh}, Griffin and {Jenness}, Tim and {Jones}, Craig K. and {Joseph}, Prajwel and {Kalmbach}, J. Bryce and {Karamehmetoglu}, Emir and {Ka{\l}uszy{\'n}ski}, Miko{\l}aj and {Kelley}, Michael S.~P. and {Kern}, Nicholas and {Kerzendorf}, Wolfgang E. and {Koch}, Eric W. and {Kulumani}, Shankar and {Lee}, Antony and {Ly}, Chun and {Ma}, Zhiyuan and {MacBride}, Conor and {Maljaars}, Jakob M. and {Muna}, Demitri and {Murphy}, N.~A. and {Norman}, Henrik and {O'Steen}, Richard and {Oman}, Kyle A. and {Pacifici}, Camilla and {Pascual}, Sergio and {Pascual-Granado}, J. and {Patil}, Rohit R. and {Perren}, Gabriel I. and {Pickering}, Timothy E. and {Rastogi}, Tanuj and {Roulston}, Benjamin R. and {Ryan}, Daniel F. and {Rykoff}, Eli S. and {Sabater}, Jose and {Sakurikar}, Parikshit and {Salgado}, Jes{\'u}s and {Sanghi}, Aniket and {Saunders}, Nicholas and {Savchenko}, Volodymyr and {Schwardt}, Ludwig and {Seifert-Eckert}, Michael and {Shih}, Albert Y. and {Jain}, Anany Shrey and {Shukla}, Gyanendra and {Sick}, Jonathan and {Simpson}, Chris and {Singanamalla}, Sudheesh and {Singer}, Leo P. and {Singhal}, Jaladh and {Sinha}, Manodeep and {Sip{\H{o}}cz}, Brigitta M. and {Spitler}, Lee R. and {Stansby}, David and {Streicher}, Ole and {{\v{S}}umak}, Jani and {Swinbank}, John D. and {Taranu}, Dan S. and {Tewary}, Nikita and {Tremblay}, Grant R. and {de Val-Borro}, Miguel and {Van Kooten}, Samuel J. and {Vasovi{\'c}}, Zlatan and {Verma}, Shresth and {de Miranda Cardoso}, Jos{\'e} Vin{\'\i}cius and {Williams}, Peter K.~G. and {Wilson}, Tom J. and {Winkel}, Benjamin and {Wood-Vasey}, W.~M. and {Xue}, Rui and {Yoachim}, Peter and {Zhang}, Chen and {Zonca}, Andrea and {Astropy Project Contributors}},
        title = "{The Astropy Project: Sustaining and Growing a Community-oriented Open-source Project and the Latest Major Release (v5.0) of the Core Package}",
      journal = {\apj},
         year = 2022,
        month = aug,
       volume = {935},
       number = {2},
          eid = {167},
        pages = {167},
          doi = {10.3847/1538-4357/ac7c74},
archivePrefix = {arXiv},
       eprint = {2206.14220},
 primaryClass = {astro-ph.IM},
       adsurl = {https://ui.adsabs.harvard.edu/abs/2022ApJ...935..167A}
}

@ARTICLE{2020SciPy-NMeth,
  author  = {Virtanen, Pauli and Gommers, Ralf and Oliphant, Travis E. and
            Haberland, Matt and Reddy, Tyler and Cournapeau, David and
            Burovski, Evgeni and Peterson, Pearu and Weckesser, Warren and
            Bright, Jonathan and {van der Walt}, St{\'e}fan J. and
            Brett, Matthew and Wilson, Joshua and Millman, K. Jarrod and
            Mayorov, Nikolay and Nelson, Andrew R. J. and Jones, Eric and
            Kern, Robert and Larson, Eric and Carey, C J and
            Polat, {\.I}lhan and Feng, Yu and Moore, Eric W. and
            {VanderPlas}, Jake and Laxalde, Denis and Perktold, Josef and
            Cimrman, Robert and Henriksen, Ian and Quintero, E. A. and
            Harris, Charles R. and Archibald, Anne M. and
            Ribeiro, Ant{\^o}nio H. and Pedregosa, Fabian and
            {van Mulbregt}, Paul and {SciPy 1.0 Contributors}},
  title   = {{{SciPy} 1.0: Fundamental Algorithms for Scientific
            Computing in Python}},
  journal = {Nature Methods},
  year    = {2020},
  volume  = {17},
  pages   = {261--272},
  adsurl  = {https://rdcu.be/b08Wh},
  doi     = {10.1038/s41592-019-0686-2},
}

@Article{Hunter:2007,
  Author    = {Hunter, J. D.},
  Title     = {Matplotlib: A 2D graphics environment},
  Journal   = {Computing in Science \& Engineering},
  Volume    = {9},
  Number    = {3},
  Pages     = {90--95},
  publisher = {IEEE COMPUTER SOC},
  doi       = {10.1109/MCSE.2007.55},
  year      = 2007
}

@software{ESOCPLDevelopmentTeam_2015,
       author = {{ESO CPL Development Team}},
 howpublished = {Astrophysics Source Code Library, record ascl:1504.003},
         year = 2015,
        month = apr,
archivePrefix = {ascl},
       eprint = {1504.003},
       adsurl = {https://ui.adsabs.harvard.edu/abs/2015ascl.soft04003E}
}

@ARTICLE{Harris_2020,
       author = {{Harris}, Charles R. and {Millman}, K. Jarrod and {van der Walt}, St{\'e}fan J. and {Gommers}, Ralf and {Virtanen}, Pauli and {Cournapeau}, David and {Wieser}, Eric and {Taylor}, Julian and {Berg}, Sebastian and {Smith}, Nathaniel J. and {Kern}, Robert and {Picus}, Matti and {Hoyer}, Stephan and {van Kerkwijk}, Marten H. and {Brett}, Matthew and {Haldane}, Allan and {del R{\'\i}o}, Jaime Fern{\'a}ndez and {Wiebe}, Mark and {Peterson}, Pearu and {G{\'e}rard-Marchant}, Pierre and {Sheppard}, Kevin and {Reddy}, Tyler and {Weckesser}, Warren and {Abbasi}, Hameer and {Gohlke}, Christoph and {Oliphant}, Travis E.},
        title = "{Array programming with NumPy}",
      journal = {\nat},
         year = 2020,
        month = sep,
       volume = {585},
       number = {7825},
        pages = {357-362},
          doi = {10.1038/s41586-020-2649-2},
archivePrefix = {arXiv},
       eprint = {2006.10256},
 primaryClass = {cs.MS},
       adsurl = {https://ui.adsabs.harvard.edu/abs/2020Natur.585..357H}
}

@article{Wesson_2004,
    author = {Wesson, R. and Liu, X.-W.},
    title = {Physical conditions in the planetary nebula NGC 6543},
    journal = {\mnras},
    volume = {351},
    number = {3},
    pages = {1026-1042},
    year = {2004},
    month = {07},
    issn = {0035-8711},
    doi = {10.1111/j.1365-2966.2004.07856.x},
    url = {https://doi.org/10.1111/j.1365-2966.2004.07856.x},
    eprint = {https://academic.oup.com/mnras/article-pdf/351/3/1026/3570037/351-3-1026.pdf},
}

@ARTICLE{Lovis2007,
       author = {{Lovis}, C. and {Pepe}, F.},
        title = "{A new list of thorium and argon spectral lines in the visible}",
      journal = {\aap},
         year = 2007,
        month = jun,
       volume = {468},
       number = {3},
        pages = {1115-1121},
          doi = {10.1051/0004-6361:20077249},
archivePrefix = {arXiv},
       eprint = {astro-ph/0703412},
 primaryClass = {astro-ph},
       adsurl = {https://ui.adsabs.harvard.edu/abs/2007A&A...468.1115L}
}

@article{Hauschildt1999,
  author = {Hauschildt, Peter H. and Baron, E.},
  title = {Numerical Solution of the Expanding Stellar Atmosphere Problem},
  journal = {Journal of Computational and Applied Mathematics},
  year = {1999},
  volume = {109},
  pages = {41--63}
}

@ARTICLE{TESSEB2022,
       author = {{Pr{\v{s}}a}, Andrej and {Kochoska}, Angela and {Conroy}, Kyle E. and {Eisner}, Nora and {Hey}, Daniel R. and {IJspeert}, Luc and {Kruse}, Ethan and {Fleming}, Scott W. and {Johnston}, Cole and {Kristiansen}, Martti H. and {LaCourse}, Daryll and {Mortensen}, Danielle and {Pepper}, Joshua and {Stassun}, Keivan G. and {Torres}, Guillermo and {Abdul-Masih}, Michael and {Chakraborty}, Joheen and {Gagliano}, Robert and {Guo}, Zhao and {Hambleton}, Kelly and {Hong}, Kyeongsoo and {Jacobs}, Thomas and {Jones}, David and {Kostov}, Veselin and {Lee}, Jae Woo and {Omohundro}, Mark and {Orosz}, Jerome A. and {Page}, Emma J. and {Powell}, Brian P. and {Rappaport}, Saul and {Reed}, Phill and {Schnittman}, Jeremy and {Schwengeler}, Hans Martin and {Shporer}, Avi and {Terentev}, Ivan A. and {Vanderburg}, Andrew and {Welsh}, William F. and {Caldwell}, Douglas A. and {Doty}, John P. and {Jenkins}, Jon M. and {Latham}, David W. and {Ricker}, George R. and {Seager}, Sara and {Schlieder}, Joshua E. and {Shiao}, Bernie and {Vanderspek}, Roland and {Winn}, Joshua N.},
        title = "{TESS Eclipsing Binary Stars. I. Short-cadence Observations of 4584 Eclipsing Binaries in Sectors 1-26}",
      journal = {\apjs},
         year = 2022,
        month = jan,
       volume = {258},
       number = {1},
          eid = {16},
        pages = {16},
          doi = {10.3847/1538-4365/ac324a},
archivePrefix = {arXiv},
       eprint = {2110.13382},
 primaryClass = {astro-ph.SR},
       adsurl = {https://ui.adsabs.harvard.edu/abs/2022ApJS..258...16P}
}

@article{Guo2025,
doi = {10.3847/1538-3881/ade710},
url = {https://doi.org/10.3847/1538-3881/ade710},
year = {2025},
month = {jul},
publisher = {The American Astronomical Society},
volume = {170},
number = {2},
pages = {115},
author = {Guo, Yani and Li, Kai and Wang, Liheng and Xia, Qiqi and Gao, Xiang and Xu, Jingran and Wang, Jingyi},
title = {The Investigation of 84 TESS Totally Eclipsing Contact Binaries},
journal = {The Astronomical Journal}
}

@misc{10.17909/t9-es7m-vw14,
  doi = {10.17909/T9-ES7M-VW14},
  url = {http://archive.stsci.edu/doi/resolve/resolve.html?doi=10.17909/t9-es7m-vw14},
  author = {Nardiello,  Domenico},
  title = {A PSF-Based Approach to TESS High Quality Data Of Stellar Clusters (PATHOS)},
  publisher = {STScI/MAST},
  year = {2019}
}

@ARTICLE{efosc2,
       author = {{Buzzoni}, B. and {Delabre}, B. and {Dekker}, H. and {Dodorico}, S. and {Enard}, D. and {Focardi}, P. and {Gustafsson}, B. and {Nees}, W. and {Paureau}, J. and {Reiss}, R.},
        title = "{The ESO Faint Object Spectrograph and Camera / EFOSC}",
      journal = {The Messenger},
         year = 1984,
        month = dec,
       volume = {38},
        pages = {9},
       adsurl = {https://ui.adsabs.harvard.edu/abs/1984Msngr..38....9B}
}

@ARTICLE{gaia,
       author = {{Gaia Collaboration} and {Vallenari}, A. and {Brown}, A.~G.~A. and {Prusti}, T. and {de Bruijne}, J.~H.~J. and {Arenou}, F. and {Babusiaux}, C. and {Biermann}, M. and {Creevey}, O.~L. and {Ducourant}, C. and {Evans}, D.~W. and {Eyer}, L. and {Guerra}, R. and {Hutton}, A. and {Jordi}, C. and {Klioner}, S.~A. and {Lammers}, U.~L. and {Lindegren}, L. and {Luri}, X. and {Mignard}, F. and {Panem}, C. and {Pourbaix}, D. and {Randich}, S. and {Sartoretti}, P. and {Soubiran}, C. and {Tanga}, P. and {Walton}, N.~A. and {Bailer-Jones}, C.~A.~L. and {Bastian}, U. and {Drimmel}, R. and {Jansen}, F. and {Katz}, D. and {Lattanzi}, M.~G. and {van Leeuwen}, F. and {Bakker}, J. and {Cacciari}, C. and {Casta{\~n}eda}, J. and {De Angeli}, F. and {Fabricius}, C. and {Fouesneau}, M. and {Fr{\'e}mat}, Y. and {Galluccio}, L. and {Guerrier}, A. and {Heiter}, U. and {Masana}, E. and {Messineo}, R. and {Mowlavi}, N. and {Nicolas}, C. and {Nienartowicz}, K. and {Pailler}, F. and {Panuzzo}, P. and {Riclet}, F. and {Roux}, W. and {Seabroke}, G.~M. and {Sordo}, R. and {Th{\'e}venin}, F. and {Gracia-Abril}, G. and {Portell}, J. and {Teyssier}, D. and {Altmann}, M. and {Andrae}, R. and {Audard}, M. and {Bellas-Velidis}, I. and {Benson}, K. and {Berthier}, J. and {Blomme}, R. and {Burgess}, P.~W. and {Busonero}, D. and {Busso}, G. and {C{\'a}novas}, H. and {Carry}, B. and {Cellino}, A. and {Cheek}, N. and {Clementini}, G. and {Damerdji}, Y. and {Davidson}, M. and {de Teodoro}, P. and {Nu{\~n}ez Campos}, M. and {Delchambre}, L. and {Dell'Oro}, A. and {Esquej}, P. and {Fern{\'a}ndez-Hern{\'a}ndez}, J. and {Fraile}, E. and {Garabato}, D. and {Garc{\'\i}a-Lario}, P. and {Gosset}, E. and {Haigron}, R. and {Halbwachs}, J.-L. and {Hambly}, N.~C. and {Harrison}, D.~L. and {Hern{\'a}ndez}, J. and {Hestroffer}, D. and {Hodgkin}, S.~T. and {Holl}, B. and {Jan{\ss}en}, K. and {Jevardat de Fombelle}, G. and {Jordan}, S. and {Krone-Martins}, A. and {Lanzafame}, A.~C. and {L{\"o}ffler}, W. and {Marchal}, O. and {Marrese}, P.~M. and {Moitinho}, A. and {Muinonen}, K. and {Osborne}, P. and {Pancino}, E. and {Pauwels}, T. and {Recio-Blanco}, A. and {Reyl{\'e}}, C. and {Riello}, M. and {Rimoldini}, L. and {Roegiers}, T. and {Rybizki}, J. and {Sarro}, L.~M. and {Siopis}, C. and {Smith}, M. and {Sozzetti}, A. and {Utrilla}, E. and {van Leeuwen}, M. and {Abbas}, U. and {{\'A}brah{\'a}m}, P. and {Abreu Aramburu}, A. and {Aerts}, C. and {Aguado}, J.~J. and {Ajaj}, M. and {Aldea-Montero}, F. and {Altavilla}, G. and {{\'A}lvarez}, M.~A. and {Alves}, J. and {Anders}, F. and {Anderson}, R.~I. and {Anglada Varela}, E. and {Antoja}, T. and {Baines}, D. and {Baker}, S.~G. and {Balaguer-N{\'u}{\~n}ez}, L. and {Balbinot}, E. and {Balog}, Z. and {Barache}, C. and {Barbato}, D. and {Barros}, M. and {Barstow}, M.~A. and {Bartolom{\'e}}, S. and {Bassilana}, J.-L. and {Bauchet}, N. and {Becciani}, U. and {Bellazzini}, M. and {Berihuete}, A. and {Bernet}, M. and {Bertone}, S. and {Bianchi}, L. and {Binnenfeld}, A. and {Blanco-Cuaresma}, S. and {Blazere}, A. and {Boch}, T. and {Bombrun}, A. and {Bossini}, D. and {Bouquillon}, S. and {Bragaglia}, A. and {Bramante}, L. and {Breedt}, E. and {Bressan}, A. and {Brouillet}, N. and {Brugaletta}, E. and {Bucciarelli}, B. and {Burlacu}, A. and {Butkevich}, A.~G. and {Buzzi}, R. and {Caffau}, E. and {Cancelliere}, R. and {Cantat-Gaudin}, T. and {Carballo}, R. and {Carlucci}, T. and {Carnerero}, M.~I. and {Carrasco}, J.~M. and {Casamiquela}, L. and {Castellani}, M. and {Castro-Ginard}, A. and {Chaoul}, L. and {Charlot}, P. and {Chemin}, L. and {Chiaramida}, V. and {Chiavassa}, A. and {Chornay}, N. and {Comoretto}, G. and {Contursi}, G. and {Cooper}, W.~J. and {Cornez}, T. and {Cowell}, S. and {Crifo}, F. and {Cropper}, M. and {Crosta}, M. and {Crowley}, C. and {Dafonte}, C. and {Dapergolas}, A. and {David}, M. and {David}, P. and {de Laverny}, P. and {De Luise}, F. and {De March}, R.},
        title = "{Gaia Data Release 3. Summary of the content and survey properties}",
      journal = {\aap},
         year = 2023,
        month = jun,
       volume = {674},
          eid = {A1},
        pages = {A1},
          doi = {10.1051/0004-6361/202243940},
archivePrefix = {arXiv},
       eprint = {2208.00211},
 primaryClass = {astro-ph.GA},
       adsurl = {https://ui.adsabs.harvard.edu/abs/2023A&A...674A...1G}
}

@article{AnthonyTwarog_2018,
doi = {10.3847/1538-3881/aaad66},
url = {https://doi.org/10.3847/1538-3881/aaad66},
year = {2018},
month = {mar},
publisher = {The American Astronomical Society},
volume = {155},
number = {3},
pages = {138},
author = {Anthony-Twarog, Barbara J. and Lee-Brown, Donald B. and Deliyannis, Constantine P. and Twarog, Bruce A.},
title = {WIYN Open Cluster Study. LXXVI. Li Evolution Among Stars of Low/Intermediate Mass: The Metal-deficient Open Cluster NGC 2506},
journal = {The Astronomical Journal}
}

@ARTICLE{2021Rain,
       author = {{Rain}, M.~J. and {Ahumada}, J.~A. and {Carraro}, G.},
        title = "{A new, Gaia-based, catalogue of blue straggler stars in open clusters}",
      journal = {\aap},
         year = 2021,
        month = jun,
       volume = {650},
          eid = {A67},
        pages = {A67},
          doi = {10.1051/0004-6361/202040072},
archivePrefix = {arXiv},
       eprint = {2103.06004},
 primaryClass = {astro-ph.SR},
       adsurl = {https://ui.adsabs.harvard.edu/abs/2021A&A...650A..67R}
}

@ARTICLE{2007Arentoft,
       author = {{Arentoft}, T. and {De Ridder}, J. and {Grundahl}, F. and {Glowienka}, L. and {Waelkens}, C. and {Dupret}, M.-A. and {Grigahc{\`e}ne}, A. and {Lefever}, K. and {Jensen}, H.~R. and {Reyniers}, M. and {Frandsen}, S. and {Kjeldsen}, H.},
        title = "{Oscillating blue stragglers, {\ensuremath{\gamma}} Doradus stars and eclipsing binaries in the open cluster NGC 2506}",
      journal = {\aap},
         year = 2007,
        month = apr,
       volume = {465},
       number = {3},
        pages = {965-979},
          doi = {10.1051/0004-6361:20066931},
archivePrefix = {arXiv},
       eprint = {astro-ph/0703111},
 primaryClass = {astro-ph},
       adsurl = {https://ui.adsabs.harvard.edu/abs/2007A&A...465..965A}
}
\bibliographystyle{aasjournalv7}

\end{document}